\documentclass[twocolumn]{aastex631}

\usepackage{amsmath}

\begin{document}

\title{Detection of an Atmospheric Outflow from the Young Hot Saturn TOI-1268b}

\author[0000-0001-7144-589X]{Jorge Pérez-González}
\affiliation{Department of Physics and Astronomy, University College London,Gower Street, WC1E 6BT London, UK }

\author[0000-0002-0371-1647]{Michael Greklek-McKeon}
\affiliation{Division of Geological and Planetary Sciences, California Institute of Technology, Pasadena, CA 91125, USA}

\author[0000-0003-2527-1475]{Shreyas Vissapragada}
\affiliation{Center for Astrophysics|Harvard $\&$ Smithsonian, 60 Garden Street, Cambridge, MA 02138, USA}

\author[0000-0001-9518-9691]{Morgan Saidel}
\affiliation{Division of Geological and Planetary Sciences, California Institute of Technology, Pasadena, CA 91125, USA}

\author[0000-0002-5375-4725]{Heather A. Knutson}
\affiliation{Division of Geological and Planetary Sciences, California Institute of Technology, Pasadena, CA 91125, USA}

\author[0000-0001-6025-6663]{Dion Linssen}
\affiliation{Anton Pannekoek Institute of Astronomy, University of Amsterdam, Science Park 904, 1098 XH Amsterdam, The Netherlands}

\author[0000-0002-9584-6476]{ Antonija Oklopčić}
\affiliation{Anton Pannekoek Institute of Astronomy, University of Amsterdam, Science Park 904, 1098 XH Amsterdam, The Netherlands}



\begin{abstract}
Photoevaporative mass-loss rates are expected to be highest when planets are young and the host star is more active, but to date there have been relatively few measurements of mass-loss rates for young gas giant exoplanets. In this study we measure the present-day atmospheric mass-loss rate of TOI-1268b, a young (110 - 380 Myr) and low density (0.71$^{+0.17}_{-0.13}$~g~cm$^{-3}$) hot Saturn located near the upper edge of the Neptune desert.  We use Palomar/WIRC to search for excess absorption in the 1083~nm helium triplet during two transits of TOI-1268b.  We find that it has a larger transit depth ($0.285_{-0.050}^{+0.048}\%$ excess) in the helium bandpass than in the TESS observations, and convert this excess absorption into a mass-loss rate by modeling the outflow as a Parker wind.  Our results indicate that this planet is losing mass at a rate of $\log \dot{M} = 10.2 \pm 0.3$~g~s$^{-1}$ and has a thermosphere temperature of 6900$^{+1800}_{-1200}$~K.  This corresponds to a predicted atmospheric lifetime much larger than 10 Gyr.  Our result suggests that photoevaporation is weak in gas giant exoplanets even at early ages.
\end{abstract}

\keywords{Exoplanet atmospheres (487) --- Narrow band photometry (1088)}


\section{Introduction} 
\label{sec:intro}

Planets on close-in orbits receive extreme amounts of high-energy radiation from their host star, which can cause their atmospheres to undergo hydrodynamic escape \citep{owen2019atmospheric}.  Young stars typically have high activity levels \citep{johnstone_2021, King_2021}, and young planets that are still radiating away residual heat from their formation have low surface gravities.  As a result, most atmospheric mass loss is thought to take place at relatively early times \citep[$<$ Gyr; e.g.,][]{Kubyshkina_2022, Ketzer_2023}. However, radiative transfer modeling of photoevaporative outflows also predicts that the stellar high-energy flux does work powering the outflow, thus modifying the mass loss efficiency \citep[e.g.,][]{MurrayClay_2009,salz2016,Caldiroli_2022}. For young planets on close-in orbits  with higher XUV fluxes the corresponding mass loss efficiency will be reduced (see equation \ref{massloss_energy}). 
\begin{equation}
    \dot{M} = \frac{\varepsilon\pi R_\mathrm{p}^3F_\mathrm{XUV}}{GM_\mathrm{p}} 
    \label{massloss_energy}
\end{equation}
Where $\dot{M}$ is the mass loss rate, $\varepsilon$ is the mass loss efficiency, $R_\mathrm{p}$ is the planet's radius, $F_\mathrm{XUV}$ is the incident XUV flux from the host star, $G$ is the gravitational constant and $M_\mathrm{p}$ is the planet's mass. Young stars also have fast and dense winds, which can strongly sculpt the planetary outflow geometry and may alter the corresponding mass-loss rates \citep[e.g.,][]{Kubyshkina_2021_apr, Kubyshkina_2021_jun, MacLeod_2022, Wang_2021, carolan_2021}. In order to understand the net effect of these competing processes, it is therefore critical to measure the mass-loss rates of planets orbiting young stars.

\begin{figure}[ht]
\centering
\includegraphics[width=0.5\textwidth]{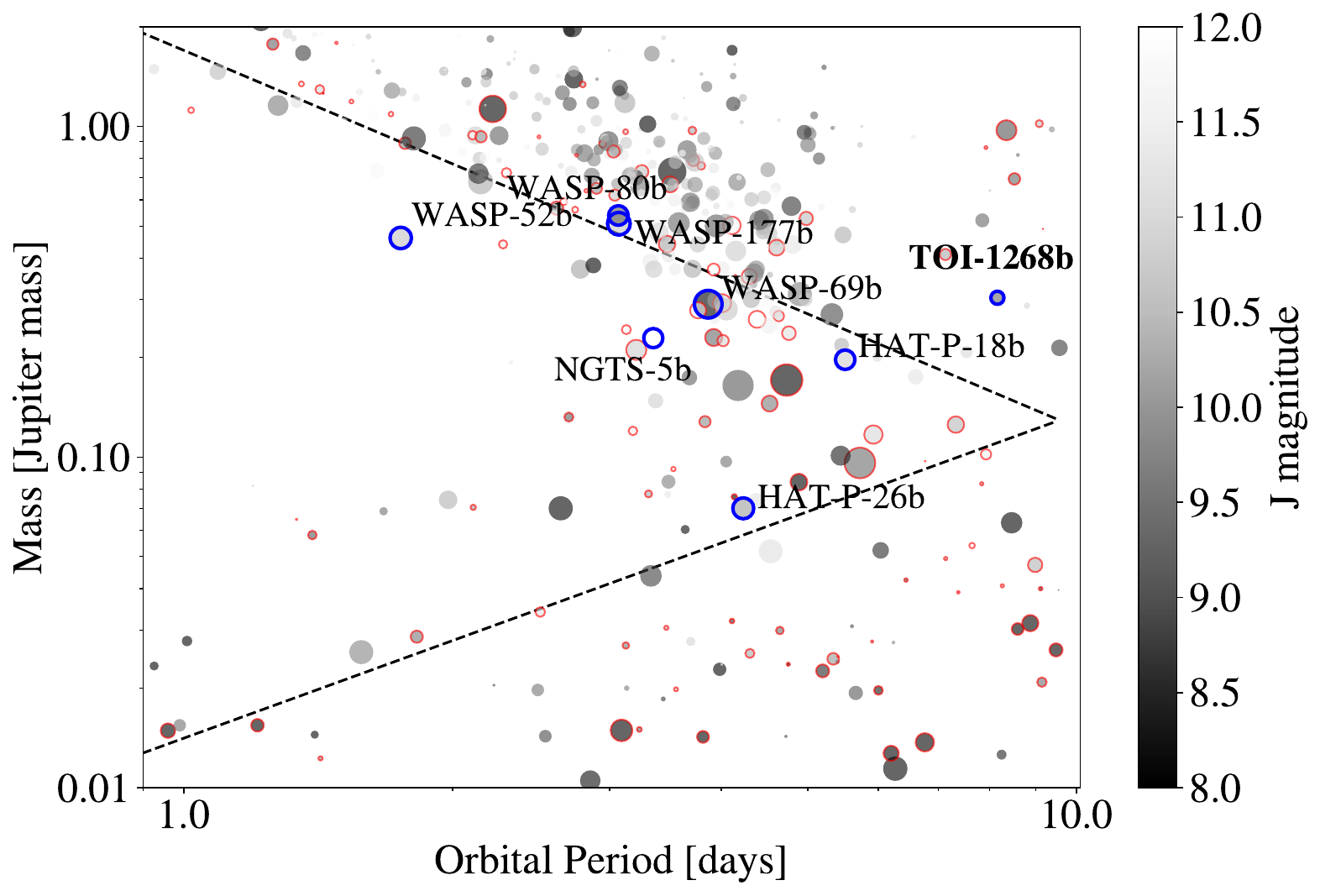}
\caption{Mass versus orbital period for the population of transiting exoplanets with fractional mass uncertainties smaller 30\%. The Neptune desert boundaries from \citet{Mazeh_2016} are overplotted as dashed lines. Planets orbiting stars with $4000$~K$<T_\mathrm{eff}<5400$~K are the most favorable targets for 1083 nm metastable helium observations; these systems are outlined in red. The seven planets with previously published Palomar/WIRC helium measurements tabulated in \citet{vissapragada2022upper} and TOI-1268b are outlined in blue. Grey shading indicates the magnitude of each star, and the point size is proportional to the logarithm of the predicted signal-to-noise ratio in the WIRC helium bandpass.  Figure adapted from \citet{vissapragada2022upper}.}
\label{fig:desert}
\end{figure}

There are currently very few observational constraints on mass-loss rates for young transiting exoplanets \citep[e.g.,][]{dosSantos_review_2022}. We can quantify the atmospheric mass-loss rates of transiting planets by measuring the excess absorption in either the Lyman $\alpha$ or the 1083~nm helium absorption triplet \citep{dosSantos_review_2022, owen2019atmospheric}. Both of these features are strongly absorbing at low pressures, resulting in a significantly deeper transit when there is an atmospheric outflow \citep[for a more detailed explanation see e.g.][]{vidal_madjar_2003, Salz_2015, spake2018,oklopvcic2018new}. In a series of recent papers, \cite{zhang2022_feb, zhang_2022_dec, zhang_2023} used a combination of Lyman $\alpha$ and metastable helium absorption to measure the mass-loss rates for a sample of five sub-Neptune-sized ($2-3$ R$_{\oplus}$) planets orbiting young stars.  However, three of these planets do not have measured masses, while the remaining two (HD 63433c and TOI-560b) have relative uncertainties greater than $20\%$ on their measured masses \citep{barragan_2021, mallorquin_2023}.  This complicates the interpretation of their atmospheric absorption signatures during the transit.  Stellar activity poses an additional challenge when measuring mass-loss rates for sub-Neptune-sized planets orbiting young stars, as the stellar Ly$\alpha$ emission and 1083~nm metastable helium absorption features can vary by amounts comparable to the white-light transit depth \citep{zhang2022_feb, zhang_2023}. 

Young transiting gas giant planets provide a promising alternative pathway to study mass loss in young planetary systems. Their masses are easier to measure than those of young transiting sub-Neptunes, while their larger planet-to-star radius ratios allow for more precise constraints on their atmospheric mass-loss rates.  Of the sample of gas giant exoplanets with published mass-loss rates, only two (WASP-52b and WASP-80b) have - ages less than a Gyr ($400^{+300}_{-200}$ Myr; \citealp{hebrand_2013} and $100^{+30}_{-20}$ Myr; \citealp{Triaud_2013} respectively). While these estimates should be interpreted with caution as angular momentum transfer between close-in Jupiters and their host stars can obfuscate gyrochronology \citep{Lanza_2010, Poppenhaeger_2014, Mancini_2017}, intriguingly, both of these planets appear to have weaker-than-predicted outflows \citep{vissapragada2022upper,Kirk_2022,Fossati_2022}.  

In this work, we utilize the 1083~nm metastable helium absorption triplet \citep{spake2018, Nortmann2018} to measure the present-day mass-loss rate 
of the Jupiter-sized \citep[$0.81\pm0.05R_\mathrm{J}$;][]{vsubjak2022toi}, Saturn-mass \citep[$0.303\pm0.026M_\mathrm{J}$;][]{vsubjak2022toi} planet TOI-1268b. This is the youngest ($110-380$ Myr) hot Saturn-mass transiting exoplanet currently known, and one of the only young transiting planets with a well-measured mass. It orbits an early K star with a mass of $0.90\pm0.13 M_\odot$ and a radius of $0.86\pm0.02 R_\odot$ \citep{dong2022neid}. With a planetary density of just 0.71$^{+0.17}_{-0.13}$~g~cm$^{-3}$ \citep{vsubjak2022toi} and a $J$-band magnitude of $9.40\pm0.02$ \citep{Skrutskie_2006}, TOI-1268b is an ideal target for helium mass loss studies. Rossiter-McLaughlin measurements also indicate that the planet's orbit is aligned with its host star \citep{dong2022neid}, which along with the system's relatively young age appears to disfavor at least some forms of high-eccentricity migration \citep{Albrecht_2022}. 

This planet is also notable because it resides at the edge of the `Neptune desert' (see Figure \ref{fig:desert}), which is a deficit of Neptune-sized planets on close-in orbits \citep{szabo2011short, beauge2012emerging, lundkvist2016hot}.  The mechanisms that create this desert are linked to the origins of these planets, which may have been formed \textit{in situ} or migrated inward to their current orbits \citep{Bailey_2018,Fortney_2021}. Several studies have suggested that the upper edge of the desert might be explained by high eccentricity migration followed by tidal disruption of planets with periastron distances that bring them too close to the star \citep{owen_lai_2018, matsakos_2016}. Other studies have suggested that photoevaporation may clear out the desert \citep{Kurokawa_2014, Thorngren_2023}, but we previously surveyed photoevaporation rates for planets near the upper edge of the desert (see Figure \ref{fig:desert}), and found that atmospheric mass loss for planets in this region  are too low to sculpt this gap \citep{vissapragada2020constraints, vissapragada2022upper,paragas2021metastable}.  One limitation of this study was the relatively old ages of the planets in our sample, which forced us to extrapolate back in time in order to calculate the cumulative mass-loss rates for the planets in our sample.  TOI-1268b is located near the upper edge of the desert, and therefore provides us with an opportunity to circumvent this limitation by probing mass loss directly and quantitatively during the period when it may be most important for planetary evolution.

In this study, we characterize the mass-loss rate of TOI-1268b for the first time. We observed one full transit and one partial transit with the the Wide-Field Infrared Camera (WIRC) on the Hale 200" Telescope on Palomar Observatory using a narrow-band filter centered on the 1083~nm helium line. In \S\ref{sec: observations} we describe our observations.  In \S\ref{sec:lightcurve} we perform a joint fit with the phased TESS light curve to better constrain the transit shape and provide a baseline white-light transit depth for comparison. In \S\ref{sec:results}, we discuss the implications of our measurement for this planet's mass loss history.

\section{Observations}
\label{sec: observations}

\subsection{Palomar/WIRC}
We observed two transits of TOI-1268b on the nights of UT 11 March 2022 and 29 April 2022.  During the first night conditions were mostly good, with thin clouds that modestly reduced the total number of counts during the second half of the night. At the start of the night the airmass value was 2.41, at the end it was 1.37 and the target passed through the zenith. Conditions during the second night were more strongly affected by clouds, which prevented us from taking data during the transit ingress. At the start of the night the airmass value was 1.32, at the end it was 1.87 and the target passed through the zenith. We additionally discarded the first nine images from the second night, as the cloud cover during these exposures was thick enough that the telescope tracking algorithm struggled to calculate the centroid of the the point spread functions (PSFs). 

We utilized a beam-shaping diffuser to reshape the PSFs of stars into a 3$\arcsec$ top hat, which helps control time-correlated systematics \citep{ 2017ApJ...848....9S,vissapragada2020constraints}.  Following the procedure described in \cite{vissapragada2020constraints} and \cite{paragas2021metastable}, we first performed the standard astronomical image calibration steps to create dark-subtracted, flat-fielded images.  We then stacked a set of nine calibrated dithered images to create a combined background frame, which we used to remove the OH telluric emission lines from each image;  these lines form bright radial arcs on the detector as a result of our narrow bandpass. We extracted photometric light curves for our target and a set of comparison stars using aperture photometry with radii ranging from 1 to 25 pixels (0$\farcs$25 to 6$\farcs$25) using the \texttt{photutils} package \citep{photutils}.  We selected the optimal aperture for each night by choosing the aperture which minimized the rms in our fit residuals \citep[0.39\% and 0.45\% for the first and second night, respectively;][]{vissapragada2020constraints}. We found optimal aperture sizes of 12 and 14 pixels (3$\arcsec$ and 3$\farcs$5) on the first and second nights, respectively and used the same 3 comparison stars for each night.

To choose the comparison stars, we considered all detected sources which had signal-to-noise ratio values larger than 50 in the first image. We then selected three stars whose extracted photometry most closely matched the target star.
\begin{figure*}[!ht]%
    \includegraphics[width=0.49\textwidth]{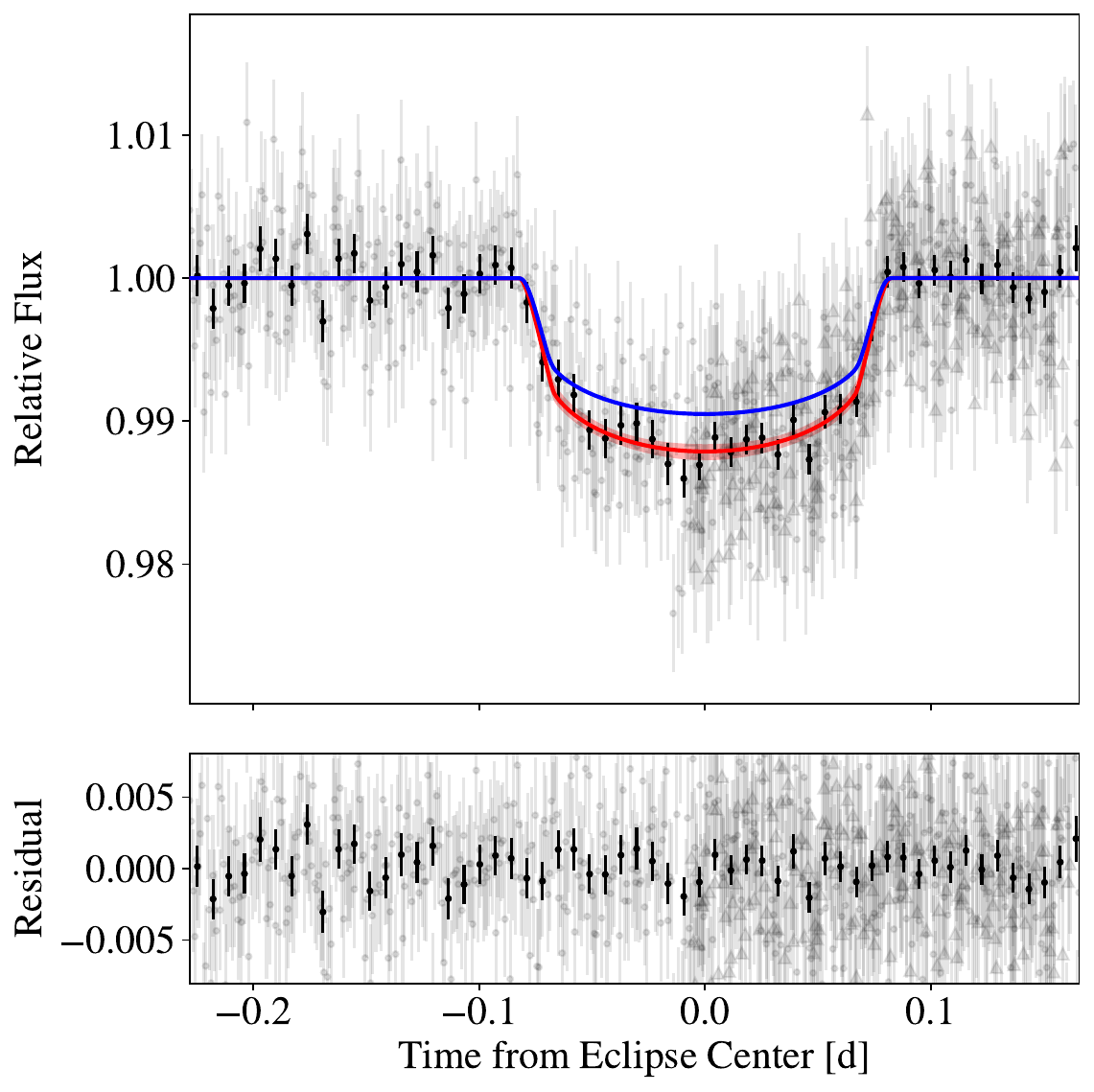}
    \includegraphics[width=0.49\textwidth]{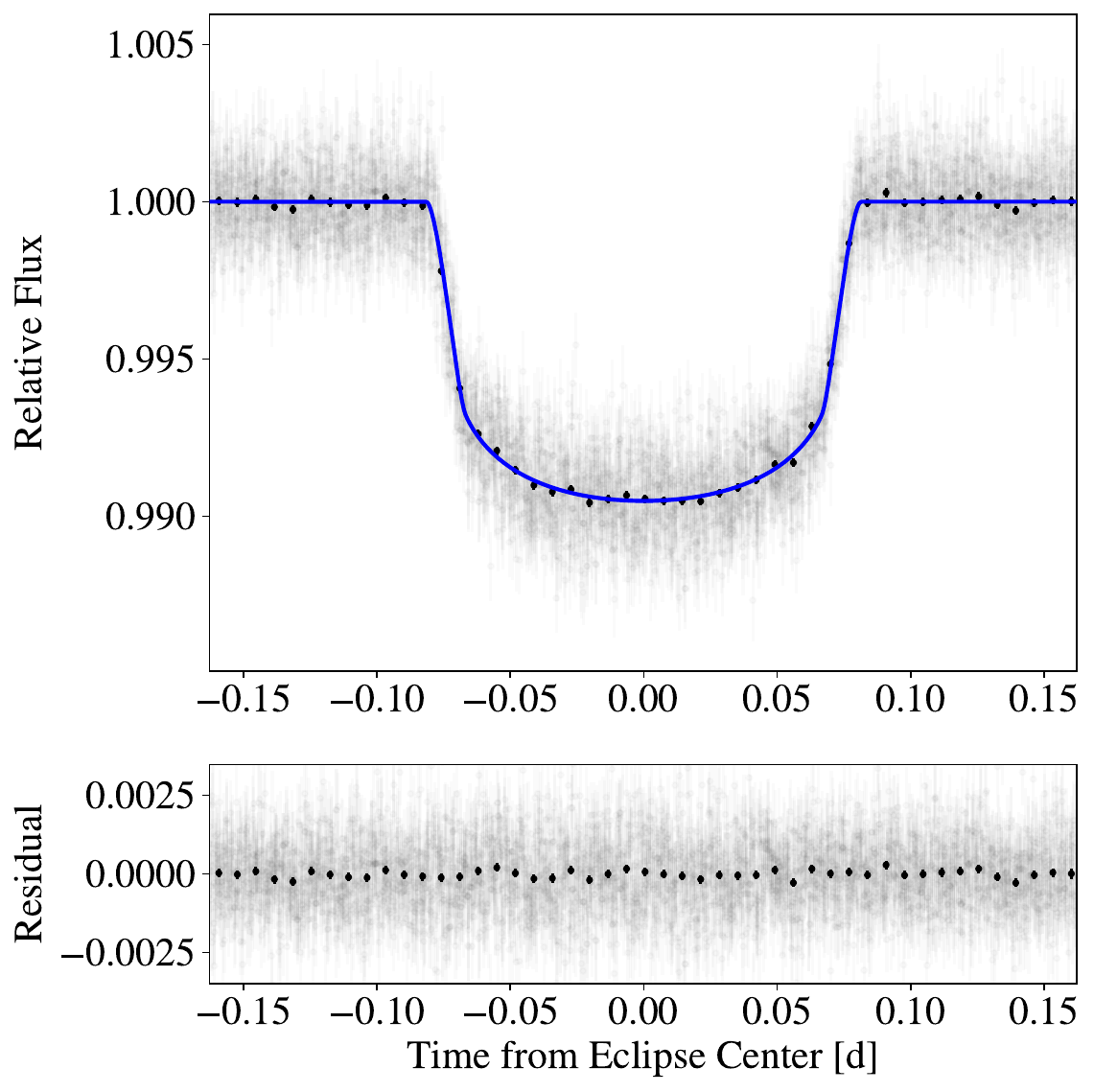}
    \caption{\underline{Left:} Combined WIRC light curve for TOI-1268b including data from both Palomar nights, with circle markers for night 1 and triangle markers for night 2. The blue curve is the best-fit transit model with the radius ratio from the \emph{TESS} bandpass, and the red curve is the best-fit transit model with the radius ratio in the WIRC helium bandpass, with the $1\sigma$ uncertainty in the transit depth shown as light red shading. \underline{Right:} \emph{TESS} data with the joint \emph{WIRC+TESS} model showed in blue.
    Unbinned photometry is shown in grey, while black points are binned to five minutes.} 
    \label{fig:double}%
\end{figure*}
\subsection{Transiting Exoplanet Survey Satellite} 

\begin{figure}[h!]
    \centering
    \includegraphics[width=0.5\textwidth]{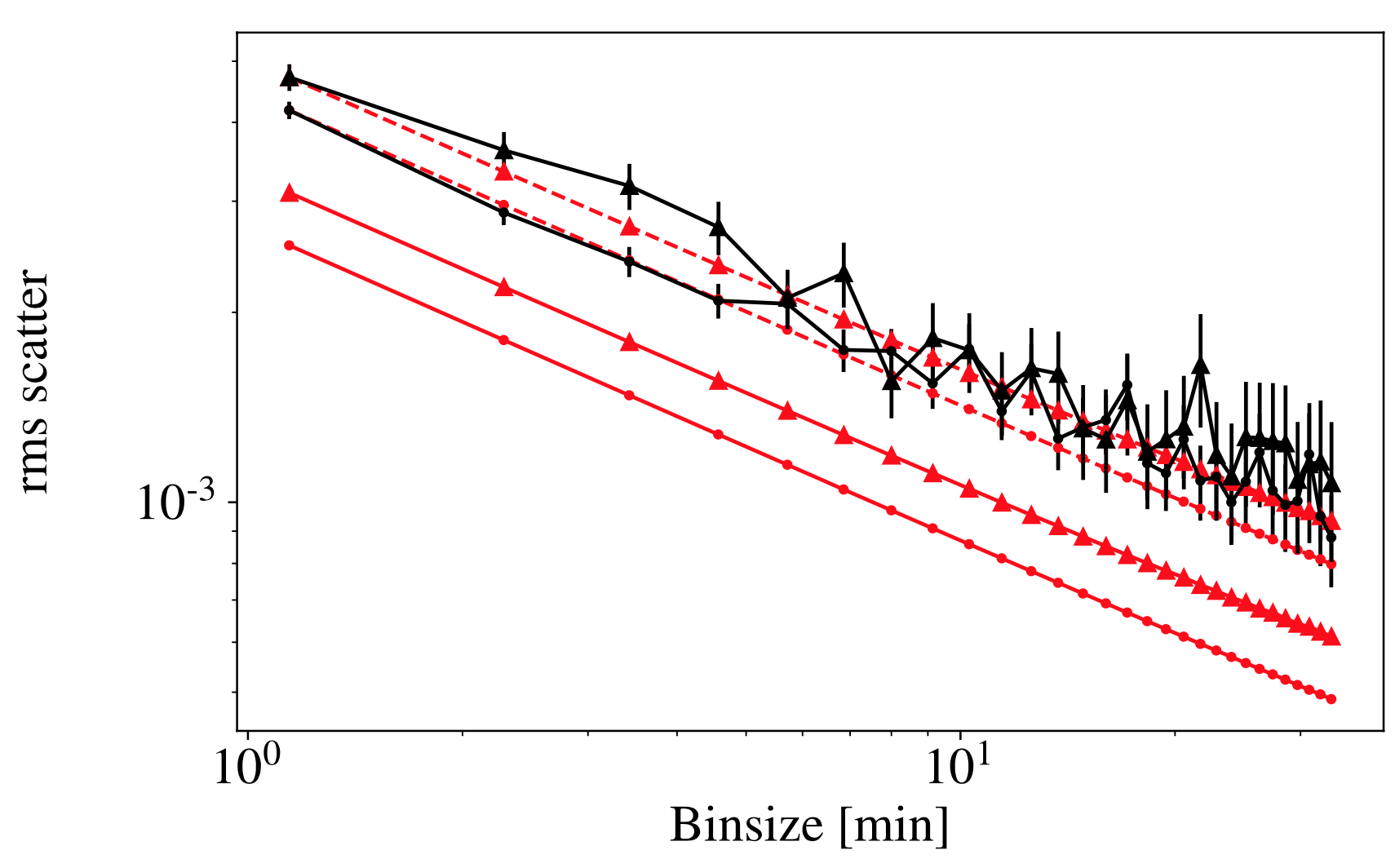}
    \caption{Allan deviation plot for the joint \emph{WIRC+TESS} fit. The black points indicate the root mean square error of the binned residuals for each light-curve, solid red line indicates the expected noise from photon noise statistics and the dashed red line indicates the photon noise scaled up the root mean square error of the binned residuals.}
    \label{fig:rms}
\end{figure}

The Transiting Exoplanet Survey Satellite (\emph{TESS}) is a space observatory that was designed to carry out a dedicated all-sky survey for transiting extrasolar planets \citep{TESS2015}. We downloaded data for TOI-1268 from \emph{TESS} sectors 15, 21, 22, 41, 48 and 49 using the Mikulski Archive for Space Telescopes (MAST). We used the Presearch Data Conditioning Simple Aperture Photometry (PDCSAP) fluxes obtained from MAST. These fluxes were extracted using the Science Processing Operations Center (SPOC) pipeline \citep{Jenkins_2016}. We then removed outliers using a 5$\sigma$ clip of the data using a moving median filter of a width of 31 data points for 5 iterations.

As discussed by \cite{vsubjak2022toi} and \cite{dong2022neid}, this relatively young star exhibits a large rotational modulation ($P=10.9\pm0.5$ days, $1-2\%$ variability) in the \emph{TESS} data due to star spots. In order to fit the transit shape, we must first remove this variability from the light curve. We calculated mid transit times for all transits using the linear ephemeris from \cite{vsubjak2022toi}. We masked out all of the transits using a 1.04 transit duration ($T_{14} = 4.001^{+0.025}_{-0.024}$~hr) wide mask ($\pm$ 0.52 transit duration from the transit center) and then divided the data into segments, where we considered any gap longer than 1 day as a division between segments. We mirrored the data at the beginning and end of each segment to eliminate edge effects on the smoothing filter \citep{Mike_2023}, and used a Gaussian kernel convolution with a kernel width of 1 transit duration to generate a smoothed trend representing the rotational modulation. We divided this trend out of our \emph{TESS} data to normalize and flatten the light curve. We also found that transit numbers 13 and 15 in the \emph{TESS} photometry had star spot crossings, which we identified by first fitting the stacked profile of all transits and then flagging transits with 5 or more consecutive points that are more than 3$\sigma$ from the best-fit stacked transit model. We removed these transits from the stacked profile so as to not bias the joint fit.

\section{Light Curve Modeling}
\label{sec:lightcurve}
We modeled each individual night of Palomar/WIRC data using the \texttt{exoplanet} package \citep{exoplanet_citation}. Our model included the quadratic limb darkening coefficients $(u_1, u_2)$, impact parameter $b$, orbital period $P$, mid-transit time $T_0$, transit depth $R_p/R_{star}$ , stellar radius $R_\star$ and semi-major axis $a/R_{star}$ as free parameters. We also included a set of additional free parameters to account for time-varying systematic noise sources, where we selected the optimal set of detrending model parameters for each night of WIRC data by minimizing the Bayesian Information Criterion (BIC, \citealp{annalsofstats}). Our final set of detrending parameters for both WIRC nights included a set of linear weights for the three comparison star light curves and the airmass of the target star. During the second night, we also added a linear function of the water proxy described in \cite{paragas2021metastable}, which tracks the time-varying telluric water absorption. We considered models with linear weights on the PSF width and the target centroid position, but found that they did not improve the quality of the fit as tracked by the BIC. 

To inform our priors for the joint WIRC+\emph{TESS} fit, we first fit the TESS light curve independently, using a model that includes the quadratic limb darkening coefficients $(u_1, u_2)$, impact parameter $b$, orbital period $P$, mid-transit time $T_0$, transit depth $R_p/R_{star}$ , stellar radius $R_\star$ and semi major axis $a/R_{star}$  as free parameters. We list the priors and posteriors for each fit parameter in Table \ref{tab:joint_fit}. 

We then fit the light curves for each WIRC night individually using the \texttt{NUTS} sampler in \texttt{PyMC3} \citep{salvatier2016probabilistic}. We ran 4 chains with 1500 tuning and 1000 sampling steps per chain and we found the Gelman-Rubin parameter $\hat{R} < 1.01$ for all sampled parameters, indicating convergence. We obtained a transit depth 
$\left(R_p/R_{star}\right)^2$ of $0.98_{-0.04}^{+0.05}\%$ for the first night and $1.0_{-0.3}^{+0.2}\%$ for the second night. These values are consistent within 1$\sigma$, indicating that the amount of helium absorption over a timescale of two months is constant at the level of our measurement errors.

Finally, we fit the two WIRC light curves simultaneously along with the phased \emph{TESS} light curve. We fit for global values of the impact parameter $b$, the orbital period $P$, the predicted mid-transit time $T_0$, the stellar radius $R_\star$ and the semi-major axis $a/R_{star}$  parameters. We allowed the transit depths  $R_p/R_{star}$ and the quadratic limb darkening coefficients $(u_1, u_2)$ to vary independently in each bandpass.  We included a \emph{TESS} error scaling parameter describing the deviation from photon noise, and we included a jitter term for each night of WIRC data log($\sigma_\text{extra})$, accounting for the discrepancy between the photon noise and the true variance in the data (i.e., $\sigma^2 = \sigma^2_\text{photon} + \sigma^2_\text{extra}$).

Using the \texttt{NUTS} sampler (running 4 chains with 1000 tuning and 1500 sampling steps per chain) we obtain the posterior probability distributions for the joint fit with $\hat{R} < 1.01$. The detrended light curve and residuals are displayed in Figure \ref{fig:double}. The Allan deviation plot for the joint fit of the WIRC and \emph{TESS} data is displayed in Figure \ref{fig:rms}. The priors used and the posteriors obtained are given in Table~\ref{tab:joint_fit}.

We measure a transit depth $(R_p/R_{star})^2$ of $1.051_\pm0.047\%$ in the helium bandpass. This is larger than the TESS transit depth of $0.803_{-0.011}^{+0.009}\%$ by $0.285_{-0.050}^{+0.048}\%\,(5.7\sigma)$. 

\begin{deluxetable*}{cccccc}[!t]
    \tablecolumns{4} 
    \tablewidth{300pt} 
    \tablecaption{Priors and posteriors for Palomar/WIRC joint and \emph{TESS}-only fits.}
    \tablehead{ \colhead{Parameter} & \colhead{{Prior (joint)}} & \colhead{Posterior (joint)} & \colhead{Prior (\emph{TESS)}}&\colhead{Posterior (\emph{TESS)}}&\colhead{Units}} 
    \startdata
       $T_{0}$ &$\mathcal{U}(1703.568,  1703.611)$&$1703.5943\pm0.0012$ & $\mathcal{U}(1703.5889,1703.5901) $ &  1703.5895$\pm$0.0002 & BTJD \\
       $T_{0, TESS}$ & $\mathcal{U}( 1711.712, 1711.782)$ & $1711.73730\pm0.00013$& - &-&BTJD\\
        $P$ &$\mathcal{N}(8.157728, 0.000005)$ &$8.1577279\pm0.0000050$ & $\mathcal{U}(8.157708,8.157748) $ &  $8.157732\pm0.000003$ & days\\
        $\frac{a}{R_\star}$ &$\mathcal{N}(19.1, 1.4) $&  $17.22_{-0.34}^{+0.16}$ & $\mathcal{N}(19.1, 1.4)$ & 17.66$^{+0.05}_{-0.04}$ & - \\
        $b$ &$\mathcal{U}(0, 1)$& $0.150_{-0.096}^{+0.107}$ & $\mathcal{U}(0,1) $ & 0.03$^{+0.02}_{-0.01}$ & - \\
        $\frac{R_p}{R_\star}_{WIRC}$ & $\mathcal{U}(0,0.2)$ & $0.1025\pm{0.0023}$ & - & - & -\\
        $\frac{R_p}{R_\star}_{TESS}$  & $\mathcal{U}(0, 0.3)$ &$0.08959_{-0.00048}^{+0.00060}$& $\mathcal{U}(0,0.2)$ &  $0.0898^{+0.0002}_{-0.0003}$& -\\
        $u_{1, WIRC}$ & $\mathcal{U}(0,1) $ & $0.28_{-0.17}^{+0.22}$& - & - & - \\
        $u_{2, WIRC}$ & $\mathcal{U}(-1,1) $ & $0.23\pm0.30$& - & -  & - \\
        $u_{1, TESS}$ & $\mathcal{U}(0,1) $ & $0.35\pm{0.06}$ & $\mathcal{U}(0,1)$ & 0.39$^{+0.04}_{-0.05}$ & - \\
        $u_{2, TESS}$ & $\mathcal{U}(-1,1) $ & $0.26_{-0.12}^{+0.13}$& $\mathcal{U}(-1,1) $& 0.12$^{+0.09}_{-0.08}$  & - \\
        \emph{TESS} error scaling & $\mathcal{U}(0, 5)$& $0.978\pm{0.012}$ &- &-&-\\
    \enddata
    \tablecomments{BTJD = BJD$\mathrm{TDB}$ - 2457000. We omitted the detrending weights for each night.}
     \label{tab:joint_fit}
\end{deluxetable*}

\section{Results and Mass-loss Modelling}
\label{sec:results}
\begin{figure}[h!]
    \centering
        \includegraphics[width = 0.495\textwidth]{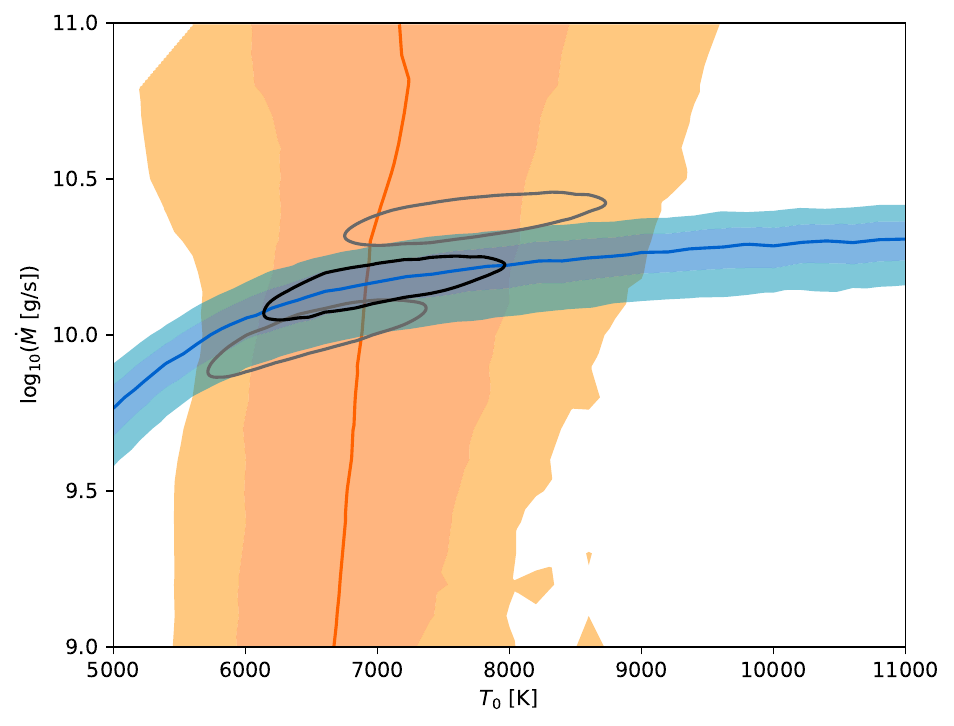}
    \caption{Combined constraint on the mass-loss rate and temperature. The orange dark and light areas are 1$\sigma$ and 2$\sigma$ contours from the temperature constrained profile. The blue dark and light areas are the best-fit models, 1$\sigma$ and 2$\sigma$ contours of the mass loss-rate profiles. The 1$\sigma$ contour of joint constraint is shown in black. The  1$\sigma$ constraints of simulations ran at three times and one third of the EUV flux are shown in gray}
    \label{fig:banana_plot}
\end{figure}

Our results indicate that TOI 1268b does indeed have an escaping atmosphere; in this section we explore the implications of our measurement for the planet's long-term evolution. Following the approach in \citet{cloudy2022}, we convert the measured excess absorption in the helium bandpass to an atmospheric mass-loss rate by modelling the outflow with a 1-dimensional Parker wind \citep{oklopvcic2018new, lampon_2020} using \texttt{Cloudy} \citep{ferland_1998,ferland_2017}. For a given mass-loss rate $\dot{M}$, a constant thermosphere temperature $T$, and a hydrogen fraction for the outflow $f_H$ that we keep fixed at 0.9, we can compute the corresponding density and velocity profiles for the atmospheric wind. Using the MUSCLES spectrum from $\epsilon$ Eridani \citep{France_2016, Loyd_2016, Youngblood_2016}, which has a similar spectral type and age to that of TOI-1268, we can then run \texttt{Cloudy} to calculate a non constant temperature structure that we compare to the assumed constant value to constrain a sensible thermospheric temperature. \texttt{Cloudy} also calculates the radially varying metastable helium density, from which we calculate the wavelength-dependent absorption signal. We average this signal over our 0.635~nm bandpass, and compare the predicted excess absorption to our measured value. We compute a grid of models spanning mass-loss rates between $10^9<\dot{M} < 10^{11}$ g$\cdot$ s$^{-1}$ and temperatures between $5000 < T <11000$ K. We then compare the predictions of this model grid to our measured excess absorption signal (see Figure \ref{fig:banana_plot}). We refer to sections 3.2 and 4.2 in \citet{cloudy2022} for additional details on the model framework. We find that our observations are best-matched by a mass-loss rate of $\log(\dot{M}) = 10.2 \pm 0.3$ g s$^{-1}$ and a corresponding thermosphere temperature of $6900^{+1800}_{-1200}$ K.

Following the approach described in section 5.3 of \citet{vissapragada2022upper}, we can convert this mass-loss rate into the corresponding mass loss efficiency $\varepsilon$ (see Eq. 2 of \citealp{vissapragada2022upper}). However, this is dependent on the incident XUV flux from the host star (see Eq. \ref{massloss_energy}).If we assume that the mass loss efficiency has remained constant over time, we can then estimate the planet's cumulative mass loss from formation to the present day.  This calculation is likely an overestimate of the true mass loss, as the mass loss efficiency should be lower at early times when the star is more active and the outflow is more highly ionized \citep{owen2012}. Integrating over the typical expression for the energy-limited mass-loss rate (see Eq. 2 of \citealp{vissapragada2022upper}) we can calculate the mass of the planet as a function of the fraction of mass lost to date, $f$ (by defining the planet's initial mass as $M_0 = M_p/f$):

\begin{equation}
    M_\mathrm{p} = \frac{1}{a}\sqrt{\Bigg(-\frac{1}{1 - \frac{1}{f^2}}\Bigg)\frac{\varepsilon R_\mathrm{p}^3E_\mathrm{XUV}}{2KG}}
    \label{eq:massdiff}
\end{equation}

where $\varepsilon$ is the mass loss efficiency, $R_p$ is the radius of the planet, $E_{\text{XUV}}$ is the integrated stellar XUV luminosity, $a$ is the semi-major axis, K is the \citet{Erkaev_2007} Roche lobe correction factor, and $G$ is the gravitational constant.

We estimated the time-integrated high energy flux $E_{\text{XUV}}$ for the star using the fast rotator track for K dwarfs from \citet{johnstone_2021} as a guide, and adopt a value of $10^{46}$ erg. Using this value, we find that TOI-1268b has only lost 0.45$\%$ of its initial mass.  This is also likely an overestimate, as the $E_{\text{XUV}}$ values reported in this study are integrated up to 1 Gyr, much older than TOI-1268b's estimated system age of 110-380 Myr \citep{vsubjak2022toi}. Furthermore, in our calculation we have adopted a fixed mass-radius relation as do \citep{owen_lai_2018}. \citet{Thorngren_2023} argue that for hot planets ($>$ 1000K), radius inflation is also time-dependent and should be integrated over alongside the luminosity in Eq. \ref{eq:massdiff}. However, at the $T_{eq} \sim 900 K$ of this planet \citep{vsubjak2022toi}, we are able to circumvent this issue, which adds some confidence to our interpretation. 

We can also look forward in time to determine whether or not TOI-1268b is expected to survive over the rest of the star's main sequence lifetime.  If we assume a constant mass-loss rate going forward (also an overestimate, as the star's activity level should decrease as it ages), we find that the time required for this planet to lose all of its mass to atmospheric escape is 584 Gyr. This is comparable to the predicted atmospheric lifetimes of other close-in gas giant planets presented in \citet{vissapragada2022upper} and \citet{Kirk_2022}. We conclude that TOI-1268b is quite stable against photoevaporation over the main-sequence lifetime of its host star. 

\section{Discussion and Conclusions}
In this study we presented the first detection of atmospheric mass loss from TOI-1268b, a young Saturn-mass planet located near the upper edge of the Neptune desert \citep{vsubjak2022toi}. We detect excess helium absorption during the transit with a significance of $5.7\sigma$, and translate this excess absorption measurement into a corresponding mass-loss rate of $\log(\dot{M}) = 10.2 \pm 0.3$ g s$^{-1}$ and a thermosphere temperature of $6900^{+1800}_{-1200}$ K. We estimate that TOI-1268b has lost less than 1\% of its total primordial envelope mass, and if it continues losing mass at the current rate, it will lose less than 1\% of its current envelope mass over the main-sequence lifetime of its host star.

This is consistent with the results of \cite{vissapragada2022upper} and other literature mass loss studies \citep{owen_lai_2018,Ionov_2018,dosSantos_review_2022}, which find that photoevaporation is negligible for gas giant planets near the upper edge of the Neptune desert.  With a density of 0.71$^{+0.17}_{-0.13}$ g cm$^{-3}$ \citep{vsubjak2022toi}, our results are also consistent with predictions from \citet{Thorngren_2023} that planets with densities higher than 0.1~g cm$^{-3}$ should be stable against photoevaporation. 

Our results indicate that TOI-1268b could have either formed in situ or arrived at its current location very early on via disk migration and still easily survive to the present day.  However, it is also possible that it has only recently migrated inward from a more distant formation location; this would further reduce its total atmospheric mass loss.  This planet appears to have a modest orbital eccentricity ($0.105^{+0.040}_{-0.039}$, \citealp{vsubjak2022toi}), potentially indicating that it arrived via high eccentricity migration and has only partially circularized. However, its orbit is also well-aligned with that of its host star \citep{dong2022neid}, which would tend to disfavor Kozai-type migration channels. In order to better constrain its past migration history, it would be helpful to obtain additional radial velocity measurements.  By extending the current radial velocity baseline, these observations could also be used to search for the presence of outer companions in the system that might have perturbed TOI-1268b's orbit. Furthermore, we calculate  the tidal realignment timescale for TOI-1268b, $\tau_R$ \citep{albrecht_2012}. We obtain a timescale of $6.35\cdot 10^{5}$ Gyr which is two orders of magnitude larger than the estimated age of the planet (110-380 Myr). Therefore realignment is highly unlikely.

Our observations of TOI-1268b represent a meaningful addition to the small sample of young transiting gas giant planets with measured atmospheric outflows.  In the future, spectroscopically-resolved observations of TOI-1268b's helium absorption signal could be used to explore how the strong winds, magnetic fields and high activity levels of young stars act to sculpt the 3D outflow geometry \citep{Schreyer2023}.  Similar observations of outflows from young sub-Neptune-sized planets suggest that they may have complex structures that are not readily matched by standard 3D models \citep{zhang_2022_dec, zhang2022_feb, zhang_2023}; it would be interesting to know if young gas giant planet outflows possess similar features.  Although we did not find any evidence of an extended comet-like tail in our data, spectroscopic observations should be more sensitive to these kinds of features \citep{Nortmann2018, spake2018,2021AJ....162..284S, 2020AJ....159..115K}. As the ongoing \emph{TESS} survey continues to identify new young transiting planet systems, it is also likely that there will be additional targets like TOI-1268b that are amenable to mass loss studies using metastable helium.  By obtaining a larger sample of mass loss measurements for young transiting gas giant planets, we can obtain a clearer view of early atmospheric mass loss processes on close-in exoplanets. 

\section*{Acknowledgements}
We are thankful to Paul Neid, Tom Barlow, Isaac Wilson and the rest of the Palomar staff for their support in obtaining these observations. JPG acknowledges support from the Summer Undergraduate Research Fellowship (SURF) at the California Institute of Technology.  SV and HAK acknowledge support for mass loss studies via the NN-EXPLORE program (2021B-0283). 

We acknowledge the use of public TESS data from pipelines at the TESS Science Office and at the TESS Science Processing Operations Center.  This paper includes data collected by the TESS mission that are publicly available from the Mikulski Archive for Space Telescopes (MAST).

The specific observations analyzed can be accessed via \citet{doi}. Funding for the TESS mission is provided by NASA's Science Mission Directorate.

This research has made use of the Exoplanet Follow-up Observation Program website, which is operated by the California Institute of Technology, under contract with the National Aeronautics and Space Administration under the Exoplanet Exploration Program.

\facilities{ADS, NASA Exoplanet Archive, Hale 200-inch, TESS.}
\software{\texttt{numpy} \citep{harris2020array},
\texttt{scipy} \citep{2020SciPy-NMeth},
\texttt{astropy} \citep{astropy:2013,astropy:2018,astropy:2022},
\texttt{matplotlib} \citep{Hunter:2007},
\texttt{theano} \citep{2016arXiv160502688full}, 
\texttt{pymc3} \citep{salvatier2016probabilistic}, 
\texttt{exoplanet} \citep{exoplanet_citation}, 
\texttt{corner} \citep{corner}, 
\texttt{cloudy} \citep{ferland_1998,ferland_2017},
\texttt{lightkurve} \citep{2018ascl.soft12013L}.}

\bibliography{sample631}{}

\begin{thebibliography}{}
\expandafter\ifx\csname natexlab\endcsname\relax\def\natexlab#1{#1}\fi
\providecommand{\url}[1]{\href{#1}{#1}}
\providecommand{\dodoi}[1]{doi:~\href{http://doi.org/#1}{\nolinkurl{#1}}}
\providecommand{\doeprint}[1]{\href{http://ascl.net/#1}{\nolinkurl{http://ascl.net/#1}}}
\providecommand{\doarXiv}[1]{\href{https://arxiv.org/abs/#1}{\nolinkurl{https://arxiv.org/abs/#1}}}

\bibitem[{Al-Rfou {et~al.}(2016)Al-Rfou, Alain, Almahairi, Angermueller, Bahdanau, Ballas, Bastien, Bayer, Belikov, Belopolsky, Bengio, Bergeron, Bergstra, Bisson, {Bleecher Snyder}, Bouchard, Boulanger-Lewandowski, Bouthillier, de~Br\'ebisson, Breuleux, Carrier, Cho, Chorowski, Christiano, Cooijmans, C\^ot\'e, C\^ot\'e, Courville, Dauphin, Delalleau, Demouth, Desjardins, Dieleman, Dinh, Ducoffe, Dumoulin, {Ebrahimi Kahou}, Erhan, Fan, Firat, Germain, Glorot, Goodfellow, Graham, Gulcehre, Hamel, Harlouchet, Heng, Hidasi, Honari, Jain, Jean, Jia, Korobov, Kulkarni, Lamb, Lamblin, Larsen, Laurent, Lee, Lefrancois, Lemieux, L\'eonard, Lin, Livezey, Lorenz, Lowin, Ma, Manzagol, Mastropietro, McGibbon, Memisevic, van Merri\"enboer, Michalski, Mirza, Orlandi, Pal, Pascanu, Pezeshki, Raffel, Renshaw, Rocklin, Romero, Roth, Sadowski, Salvatier, Savard, Schl\"uter, Schulman, Schwartz, Serban, Serdyuk, Shabanian, Simon, Spieckermann, Subramanyam, Sygnowski, Tanguay, van Tulder, Turian, Urban, Vincent, Visin, de~Vries,
  Warde-Farley, Webb, Willson, Xu, Xue, Yao, Zhang, \& Zhang}]{2016arXiv160502688full}
Al-Rfou, R., Alain, G., Almahairi, A., {et~al.} 2016, arXiv e-prints, abs/1605.02688.
\newblock \url{http://arxiv.org/abs/1605.02688}

\bibitem[{{Albrecht} {et~al.}(2012){Albrecht}, {Winn}, {Johnson}, {Howard}, {Marcy}, {Butler}, {Arriagada}, {Crane}, {Shectman}, {Thompson}, {Hirano}, {Bakos}, \& {Hartman}}]{albrecht_2012}
{Albrecht}, S., {Winn}, J.~N., {Johnson}, J.~A., {et~al.} 2012, \apj, 757, 18, \dodoi{10.1088/0004-637X/757/1/18}

\bibitem[{{Albrecht} {et~al.}(2022){Albrecht}, {Dawson}, \& {Winn}}]{Albrecht_2022}
{Albrecht}, S.~H., {Dawson}, R.~I., \& {Winn}, J.~N. 2022, \pasp, 134, 082001, \dodoi{10.1088/1538-3873/ac6c09}

\bibitem[{{Astropy Collaboration} {et~al.}(2013){Astropy Collaboration}, {Robitaille}, {Tollerud}, {Greenfield}, {Droettboom}, {Bray}, {Aldcroft}, {Davis}, {Ginsburg}, {Price-Whelan}, {Kerzendorf}, {Conley}, {Crighton}, {Barbary}, {Muna}, {Ferguson}, {Grollier}, {Parikh}, {Nair}, {Unther}, {Deil}, {Woillez}, {Conseil}, {Kramer}, {Turner}, {Singer}, {Fox}, {Weaver}, {Zabalza}, {Edwards}, {Azalee Bostroem}, {Burke}, {Casey}, {Crawford}, {Dencheva}, {Ely}, {Jenness}, {Labrie}, {Lim}, {Pierfederici}, {Pontzen}, {Ptak}, {Refsdal}, {Servillat}, \& {Streicher}}]{astropy:2013}
{Astropy Collaboration}, {Robitaille}, T.~P., {Tollerud}, E.~J., {et~al.} 2013, \aap, 558, A33, \dodoi{10.1051/0004-6361/201322068}

\bibitem[{{Astropy Collaboration} {et~al.}(2018){Astropy Collaboration}, {Price-Whelan}, {Sip{\H{o}}cz}, {G{\"u}nther}, {Lim}, {Crawford}, {Conseil}, {Shupe}, {Craig}, {Dencheva}, {Ginsburg}, {Vand erPlas}, {Bradley}, {P{\'e}rez-Su{\'a}rez}, {de Val-Borro}, {Aldcroft}, {Cruz}, {Robitaille}, {Tollerud}, {Ardelean}, {Babej}, {Bach}, {Bachetti}, {Bakanov}, {Bamford}, {Barentsen}, {Barmby}, {Baumbach}, {Berry}, {Biscani}, {Boquien}, {Bostroem}, {Bouma}, {Brammer}, {Bray}, {Breytenbach}, {Buddelmeijer}, {Burke}, {Calderone}, {Cano Rodr{\'\i}guez}, {Cara}, {Cardoso}, {Cheedella}, {Copin}, {Corrales}, {Crichton}, {D'Avella}, {Deil}, {Depagne}, {Dietrich}, {Donath}, {Droettboom}, {Earl}, {Erben}, {Fabbro}, {Ferreira}, {Finethy}, {Fox}, {Garrison}, {Gibbons}, {Goldstein}, {Gommers}, {Greco}, {Greenfield}, {Groener}, {Grollier}, {Hagen}, {Hirst}, {Homeier}, {Horton}, {Hosseinzadeh}, {Hu}, {Hunkeler}, {Ivezi{\'c}}, {Jain}, {Jenness}, {Kanarek}, {Kendrew}, {Kern}, {Kerzendorf}, {Khvalko}, {King}, {Kirkby}, {Kulkarni},
  {Kumar}, {Lee}, {Lenz}, {Littlefair}, {Ma}, {Macleod}, {Mastropietro}, {McCully}, {Montagnac}, {Morris}, {Mueller}, {Mumford}, {Muna}, {Murphy}, {Nelson}, {Nguyen}, {Ninan}, {N{\"o}the}, {Ogaz}, {Oh}, {Parejko}, {Parley}, {Pascual}, {Patil}, {Patil}, {Plunkett}, {Prochaska}, {Rastogi}, {Reddy Janga}, {Sabater}, {Sakurikar}, {Seifert}, {Sherbert}, {Sherwood-Taylor}, {Shih}, {Sick}, {Silbiger}, {Singanamalla}, {Singer}, {Sladen}, {Sooley}, {Sornarajah}, {Streicher}, {Teuben}, {Thomas}, {Tremblay}, {Turner}, {Terr{\'o}n}, {van Kerkwijk}, {de la Vega}, {Watkins}, {Weaver}, {Whitmore}, {Woillez}, {Zabalza}, \& {Astropy Contributors}}]{astropy:2018}
{Astropy Collaboration}, {Price-Whelan}, A.~M., {Sip{\H{o}}cz}, B.~M., {et~al.} 2018, \aj, 156, 123, \dodoi{10.3847/1538-3881/aabc4f}

\bibitem[{{Astropy Collaboration} {et~al.}(2022){Astropy Collaboration}, {Price-Whelan}, {Lim}, {Earl}, {Starkman}, {Bradley}, {Shupe}, {Patil}, {Corrales}, {Brasseur}, {N{"o}the}, {Donath}, {Tollerud}, {Morris}, {Ginsburg}, {Vaher}, {Weaver}, {Tocknell}, {Jamieson}, {van Kerkwijk}, {Robitaille}, {Merry}, {Bachetti}, {G{"u}nther}, {Aldcroft}, {Alvarado-Montes}, {Archibald}, {B{'o}di}, {Bapat}, {Barentsen}, {Baz{'a}n}, {Biswas}, {Boquien}, {Burke}, {Cara}, {Cara}, {Conroy}, {Conseil}, {Craig}, {Cross}, {Cruz}, {D'Eugenio}, {Dencheva}, {Devillepoix}, {Dietrich}, {Eigenbrot}, {Erben}, {Ferreira}, {Foreman-Mackey}, {Fox}, {Freij}, {Garg}, {Geda}, {Glattly}, {Gondhalekar}, {Gordon}, {Grant}, {Greenfield}, {Groener}, {Guest}, {Gurovich}, {Handberg}, {Hart}, {Hatfield-Dodds}, {Homeier}, {Hosseinzadeh}, {Jenness}, {Jones}, {Joseph}, {Kalmbach}, {Karamehmetoglu}, {Ka{l}uszy{'n}ski}, {Kelley}, {Kern}, {Kerzendorf}, {Koch}, {Kulumani}, {Lee}, {Ly}, {Ma}, {MacBride}, {Maljaars}, {Muna}, {Murphy}, {Norman}, {O'Steen},
  {Oman}, {Pacifici}, {Pascual}, {Pascual-Granado}, {Patil}, {Perren}, {Pickering}, {Rastogi}, {Roulston}, {Ryan}, {Rykoff}, {Sabater}, {Sakurikar}, {Salgado}, {Sanghi}, {Saunders}, {Savchenko}, {Schwardt}, {Seifert-Eckert}, {Shih}, {Jain}, {Shukla}, {Sick}, {Simpson}, {Singanamalla}, {Singer}, {Singhal}, {Sinha}, {Sip{H{o}}cz}, {Spitler}, {Stansby}, {Streicher}, {{{S}}umak}, {Swinbank}, {Taranu}, {Tewary}, {Tremblay}, {Val-Borro}, {Van Kooten}, {Vasovi{'c}}, {Verma}, {de Miranda Cardoso}, {Williams}, {Wilson}, {Winkel}, {Wood-Vasey}, {Xue}, {Yoachim}, {Zhang}, {Zonca}, \& {Astropy Project Contributors}}]{astropy:2022}
{Astropy Collaboration}, {Price-Whelan}, A.~M., {Lim}, P.~L., {et~al.} 2022, apj, 935, 167, \dodoi{10.3847/1538-4357/ac7c74}

\bibitem[{{Bailey} \& {Batygin}(2018)}]{Bailey_2018}
{Bailey}, E., \& {Batygin}, K. 2018, \apjl, 866, L2, \dodoi{10.3847/2041-8213/aade90}

\bibitem[{{Barrag{\'a}n} {et~al.}(2021){Barrag{\'a}n}, {Aigrain}, {Gillen}, \& {Guti{\'e}rrez-Canales}}]{barragan_2021}
{Barrag{\'a}n}, O., {Aigrain}, S., {Gillen}, E., \& {Guti{\'e}rrez-Canales}, F. 2021, Research Notes of the American Astronomical Society, 5, 51, \dodoi{10.3847/2515-5172/abef70}

\bibitem[{Beauge \& Nesvorn{\`y}(2012)}]{beauge2012emerging}
Beauge, C., \& Nesvorn{\`y}, D. 2012, The Astrophysical Journal, 763, 12

\bibitem[{Bradley {et~al.}(2022)Bradley, Sipőcz, Robitaille, Tollerud, Vinícius, Deil, Barbary, Wilson, Busko, Donath, Günther, Cara, Lim, Meßlinger, Conseil, Bostroem, Droettboom, Bray, Bratholm, Barentsen, Craig, Rathi, Pascual, Perren, Georgiev, de~Val-Borro, Kerzendorf, Bach, Quint, \& Souchereau}]{photutils}
Bradley, L., Sipőcz, B., Robitaille, T., {et~al.} 2022, astropy/photutils: 1.5.0, 1.5.0,  Zenodo, \dodoi{10.5281/zenodo.6825092}

\bibitem[{{Caldiroli} {et~al.}(2022){Caldiroli}, {Haardt}, {Gallo}, {Spinelli}, {Malsky}, \& {Rauscher}}]{Caldiroli_2022}
{Caldiroli}, A., {Haardt}, F., {Gallo}, E., {et~al.} 2022, \aap, 663, A122, \dodoi{10.1051/0004-6361/202142763}

\bibitem[{{Carolan} {et~al.}(2021){Carolan}, {Vidotto}, {Villarreal D'Angelo}, \& {Hazra}}]{carolan_2021}
{Carolan}, S., {Vidotto}, A.~A., {Villarreal D'Angelo}, C., \& {Hazra}, G. 2021, \mnras, 500, 3382, \dodoi{10.1093/mnras/staa3431}

\bibitem[{Dong {et~al.}(2022)Dong, Huang, Zhou, Dawson, Stef{\'a}nsson, Bender, Blake, Ford, Halverson, Kanodia, {et~al.}}]{dong2022neid}
Dong, J., Huang, C.~X., Zhou, G., {et~al.} 2022, The Astrophysical Journal Letters, 926, L7

\bibitem[{{Dos Santos}(2022)}]{dosSantos_review_2022}
{Dos Santos}, L.~A. 2022, arXiv e-prints, arXiv:2211.16243, \dodoi{10.48550/arXiv.2211.16243}

\bibitem[{{Erkaev} {et~al.}(2007){Erkaev}, {Kulikov}, {Lammer}, {Selsis}, {Langmayr}, {Jaritz}, \& {Biernat}}]{Erkaev_2007}
{Erkaev}, N.~V., {Kulikov}, Y.~N., {Lammer}, H., {et~al.} 2007, \aap, 472, 329, \dodoi{10.1051/0004-6361:20066929}

\bibitem[{{Ferland} {et~al.}(1998){Ferland}, {Korista}, {Verner}, {Ferguson}, {Kingdon}, \& {Verner}}]{ferland_1998}
{Ferland}, G.~J., {Korista}, K.~T., {Verner}, D.~A., {et~al.} 1998, \pasp, 110, 761, \dodoi{10.1086/316190}

\bibitem[{{Ferland} {et~al.}(2017){Ferland}, {Chatzikos}, {Guzm{\'a}n}, {Lykins}, {van Hoof}, {Williams}, {Abel}, {Badnell}, {Keenan}, {Porter}, \& {Stancil}}]{ferland_2017}
{Ferland}, G.~J., {Chatzikos}, M., {Guzm{\'a}n}, F., {et~al.} 2017, \rmxaa, 53, 385, \dodoi{10.48550/arXiv.1705.10877}

\bibitem[{Foreman-Mackey(2016)}]{corner}
Foreman-Mackey, D. 2016, The Journal of Open Source Software, 1, 24, \dodoi{10.21105/joss.00024}

\bibitem[{{Foreman-Mackey} {et~al.}(2021){Foreman-Mackey}, {Luger}, {Agol}, {Barclay}, {Bouma}, {Brandt}, {Czekala}, {David}, {Dong}, {Gilbert}, {Gordon}, {Hedges}, {Hey}, {Morris}, {Price-Whelan}, \& {Savel}}]{exoplanet_citation}
{Foreman-Mackey}, D., {Luger}, R., {Agol}, E., {et~al.} 2021, {exoplanet: Gradient-based probabilistic inference for exoplanet data \& other astronomical time series}, 0.5.1, Zenodo,  Zenodo, \dodoi{10.5281/zenodo.5834934}

\bibitem[{{Fortney} {et~al.}(2021){Fortney}, {Dawson}, \& {Komacek}}]{Fortney_2021}
{Fortney}, J.~J., {Dawson}, R.~I., \& {Komacek}, T.~D. 2021, Journal of Geophysical Research (Planets), 126, e06629, \dodoi{10.1029/2020JE006629}

\bibitem[{{Fossati} {et~al.}(2022){Fossati}, {Guilluy}, {Shaikhislamov}, {Carleo}, {Borsa}, {Bonomo}, {Giacobbe}, {Rainer}, {Cecchi-Pestellini}, {Khodachenko}, {Efimov}, {Rumenskikh}, {Miroshnichenko}, {Berezutsky}, {Nascimbeni}, {Brogi}, {Lanza}, {Mancini}, {Affer}, {Benatti}, {Biazzo}, {Bignamini}, {Carosati}, {Claudi}, {Cosentino}, {Covino}, {Desidera}, {Fiorenzano}, {Harutyunyan}, {Maggio}, {Malavolta}, {Maldonado}, {Micela}, {Molinari}, {Pagano}, {Pedani}, {Piotto}, {Poretti}, {Scandariato}, {Sozzetti}, \& {Stoev}}]{Fossati_2022}
{Fossati}, L., {Guilluy}, G., {Shaikhislamov}, I.~F., {et~al.} 2022, \aap, 658, A136, \dodoi{10.1051/0004-6361/202142336}

\bibitem[{{France} {et~al.}(2016){France}, {Loyd}, {Youngblood}, {Brown}, {Schneider}, {Hawley}, {Froning}, {Linsky}, {Roberge}, {Buccino}, {Davenport}, {Fontenla}, {Kaltenegger}, {Kowalski}, {Mauas}, {Miguel}, {Redfield}, {Rugheimer}, {Tian}, {Vieytes}, {Walkowicz}, \& {Weisenburger}}]{France_2016}
{France}, K., {Loyd}, R.~O.~P., {Youngblood}, A., {et~al.} 2016, \apj, 820, 89, \dodoi{10.3847/0004-637X/820/2/89}

\bibitem[{{Greklek-McKeon} {et~al.}(2023){Greklek-McKeon}, {Knutson}, {Vissapragada}, {Jontof-Hutter}, {Chachan}, {Thorngren}, \& {Vasisht}}]{Mike_2023}
{Greklek-McKeon}, M., {Knutson}, H.~A., {Vissapragada}, S., {et~al.} 2023, \aj, 165, 48, \dodoi{10.3847/1538-3881/ac8553}

\bibitem[{Harris {et~al.}(2020)Harris, Millman, van~der Walt, Gommers, Virtanen, Cournapeau, Wieser, Taylor, Berg, Smith, Kern, Picus, Hoyer, van Kerkwijk, Brett, Haldane, del R{\'{i}}o, Wiebe, Peterson, G{\'{e}}rard-Marchant, Sheppard, Reddy, Weckesser, Abbasi, Gohlke, \& Oliphant}]{harris2020array}
Harris, C.~R., Millman, K.~J., van~der Walt, S.~J., {et~al.} 2020, Nature, 585, 357, \dodoi{10.1038/s41586-020-2649-2}

\bibitem[{{H{\'e}brard} {et~al.}(2013){H{\'e}brard}, {Collier Cameron}, {Brown}, {D{\'\i}az}, {Faedi}, {Smalley}, {Anderson}, {Armstrong}, {Barros}, {Bento}, {Bouchy}, {Doyle}, {Enoch}, {G{\'o}mez Maqueo Chew}, {H{\'e}brard}, {Hellier}, {Lendl}, {Lister}, {Maxted}, {McCormac}, {Moutou}, {Pollacco}, {Queloz}, {Santerne}, {Skillen}, {Southworth}, {Tregloan-Reed}, {Triaud}, {Udry}, {Vanhuysse}, {Watson}, {West}, \& {Wheatley}}]{hebrand_2013}
{H{\'e}brard}, G., {Collier Cameron}, A., {Brown}, D.~J.~A., {et~al.} 2013, \aap, 549, A134, \dodoi{10.1051/0004-6361/201220363}

\bibitem[{Hunter(2007)}]{Hunter:2007}
Hunter, J.~D. 2007, Computing in Science \& Engineering, 9, 90, \dodoi{10.1109/MCSE.2007.55}

\bibitem[{{Ionov} {et~al.}(2018){Ionov}, {Pavlyuchenkov}, \& {Shematovich}}]{Ionov_2018}
{Ionov}, D.~E., {Pavlyuchenkov}, Y.~N., \& {Shematovich}, V.~I. 2018, \mnras, 476, 5639, \dodoi{10.1093/mnras/sty626}

\bibitem[{{Jenkins} {et~al.}(2016){Jenkins}, {Twicken}, {McCauliff}, {Campbell}, {Sanderfer}, {Lung}, {Mansouri-Samani}, {Girouard}, {Tenenbaum}, {Klaus}, {Smith}, {Caldwell}, {Chacon}, {Henze}, {Heiges}, {Latham}, {Morgan}, {Swade}, {Rinehart}, \& {Vanderspek}}]{Jenkins_2016}
{Jenkins}, J.~M., {Twicken}, J.~D., {McCauliff}, S., {et~al.} 2016, in Society of Photo-Optical Instrumentation Engineers (SPIE) Conference Series, Vol. 9913, Software and Cyberinfrastructure for Astronomy IV, ed. G.~{Chiozzi} \& J.~C. {Guzman}, 99133E, \dodoi{10.1117/12.2233418}

\bibitem[{{Johnstone} {et~al.}(2021){Johnstone}, {Bartel}, \& {G{\"u}del}}]{johnstone_2021}
{Johnstone}, C.~P., {Bartel}, M., \& {G{\"u}del}, M. 2021, \aap, 649, A96, \dodoi{10.1051/0004-6361/202038407}

\bibitem[{{Ketzer} \& {Poppenhaeger}(2023)}]{Ketzer_2023}
{Ketzer}, L., \& {Poppenhaeger}, K. 2023, \mnras, 518, 1683, \dodoi{10.1093/mnras/stac2643}

\bibitem[{{King} \& {Wheatley}(2021)}]{King_2021}
{King}, G.~W., \& {Wheatley}, P.~J. 2021, \mnras, 501, L28, \dodoi{10.1093/mnrasl/slaa186}

\bibitem[{{Kirk} {et~al.}(2020){Kirk}, {Alam}, {L{\'o}pez-Morales}, \& {Zeng}}]{2020AJ....159..115K}
{Kirk}, J., {Alam}, M.~K., {L{\'o}pez-Morales}, M., \& {Zeng}, L. 2020, \aj, 159, 115, \dodoi{10.3847/1538-3881/ab6e66}

\bibitem[{{Kirk} {et~al.}(2022){Kirk}, {Dos Santos}, {L{\'o}pez-Morales}, {Alam}, {Oklop{\v{c}}i{\'c}}, {MacLeod}, {Zeng}, \& {Zhou}}]{Kirk_2022}
{Kirk}, J., {Dos Santos}, L.~A., {L{\'o}pez-Morales}, M., {et~al.} 2022, \aj, 164, 24, \dodoi{10.3847/1538-3881/ac722f}

\bibitem[{{Kubyshkina} \& {Fossati}(2022)}]{Kubyshkina_2022}
{Kubyshkina}, D., \& {Fossati}, L. 2022, \aap, 668, A178, \dodoi{10.1051/0004-6361/202244916}

\bibitem[{{Kubyshkina} \& {Vidotto}(2021)}]{Kubyshkina_2021_jun}
{Kubyshkina}, D., \& {Vidotto}, A.~A. 2021, \mnras, 504, 2034, \dodoi{10.1093/mnras/stab897}

\bibitem[{{Kubyshkina} \& {Fossati}(2021)}]{Kubyshkina_2021_apr}
{Kubyshkina}, D.~I., \& {Fossati}, L. 2021, Research Notes of the American Astronomical Society, 5, 74, \dodoi{10.3847/2515-5172/abf498}

\bibitem[{{Kurokawa} \& {Nakamoto}(2014)}]{Kurokawa_2014}
{Kurokawa}, H., \& {Nakamoto}, T. 2014, \apj, 783, 54, \dodoi{10.1088/0004-637X/783/1/54}

\bibitem[{{Lamp{\'o}n} {et~al.}(2020){Lamp{\'o}n}, {L{\'o}pez-Puertas}, {Lara}, {S{\'a}nchez-L{\'o}pez}, {Salz}, {Czesla}, {Sanz-Forcada}, {Molaverdikhani}, {Alonso-Floriano}, {Nortmann}, {Caballero}, {Bauer}, {Pall{\'e}}, {Montes}, {Quirrenbach}, {Nagel}, {Ribas}, {Reiners}, \& {Amado}}]{lampon_2020}
{Lamp{\'o}n}, M., {L{\'o}pez-Puertas}, M., {Lara}, L.~M., {et~al.} 2020, \aap, 636, A13, \dodoi{10.1051/0004-6361/201937175}

\bibitem[{{Lanza}(2010)}]{Lanza_2010}
{Lanza}, A.~F. 2010, \aap, 512, A77, \dodoi{10.1051/0004-6361/200912789}

\bibitem[{{Lightkurve Collaboration} {et~al.}(2018){Lightkurve Collaboration}, {Cardoso}, {Hedges}, {Gully-Santiago}, {Saunders}, {Cody}, {Barclay}, {Hall}, {Sagear}, {Turtelboom}, {Zhang}, {Tzanidakis}, {Mighell}, {Coughlin}, {Bell}, {Berta-Thompson}, {Williams}, {Dotson}, \& {Barentsen}}]{2018ascl.soft12013L}
{Lightkurve Collaboration}, {Cardoso}, J.~V.~d.~M., {Hedges}, C., {et~al.} 2018, {Lightkurve: Kepler and TESS time series analysis in Python}, Astrophysics Source Code Library.
\newblock \doeprint{1812.013}

\bibitem[{{Linssen} {et~al.}(2022){Linssen}, {Oklop{\v{c}}i{\'c}}, \& {MacLeod}}]{cloudy2022}
{Linssen}, D.~C., {Oklop{\v{c}}i{\'c}}, A., \& {MacLeod}, M. 2022, \aap, 667, A54, \dodoi{10.1051/0004-6361/202243830}

\bibitem[{{Loyd} {et~al.}(2016){Loyd}, {France}, {Youngblood}, {Schneider}, {Brown}, {Hu}, {Linsky}, {Froning}, {Redfield}, {Rugheimer}, \& {Tian}}]{Loyd_2016}
{Loyd}, R.~O.~P., {France}, K., {Youngblood}, A., {et~al.} 2016, \apj, 824, 102, \dodoi{10.3847/0004-637X/824/2/102}

\bibitem[{Lundkvist {et~al.}(2016)Lundkvist, Kjeldsen, Albrecht, Davies, Basu, Huber, Justesen, Karoff, Silva~Aguirre, Van~Eylen, {et~al.}}]{lundkvist2016hot}
Lundkvist, M., Kjeldsen, H., Albrecht, S., {et~al.} 2016, Nature Communications, 7, 1

\bibitem[{{MacLeod} \& {Oklop{\v{c}}i{\'c}}(2022)}]{MacLeod_2022}
{MacLeod}, M., \& {Oklop{\v{c}}i{\'c}}, A. 2022, \apj, 926, 226, \dodoi{10.3847/1538-4357/ac46ce}

\bibitem[{{Mallorqu{\'\i}n} {et~al.}(2023){Mallorqu{\'\i}n}, {B{\'e}jar}, {Lodieu}, {Zapatero Osorio}, {Tabernero}, {Su{\'a}rez Mascare{\~n}o}, {Zechmeister}, {Luque}, {Pall{\'e}}, \& {Montes}}]{mallorquin_2023}
{Mallorqu{\'\i}n}, M., {B{\'e}jar}, V.~J.~S., {Lodieu}, N., {et~al.} 2023, \aap, 671, A163, \dodoi{10.1051/0004-6361/202245397}

\bibitem[{{Mancini} {et~al.}(2017){Mancini}, {Southworth}, {Raia}, {Tregloan-Reed}, {Molli{\`e}re}, {Bozza}, {Bretton}, {Bruni}, {Ciceri}, {D'Ago}, {Dominik}, {Hinse}, {Hundertmark}, {J{\o}rgensen}, {Korhonen}, {Rabus}, {Rahvar}, {Starkey}, {Calchi Novati}, {Figuera Jaimes}, {Henning}, {Juncher}, {Haugb{\o}lle}, {Kains}, {Popovas}, {Schmidt}, {Skottfelt}, {Snodgrass}, {Surdej}, \& {Wertz}}]{Mancini_2017}
{Mancini}, L., {Southworth}, J., {Raia}, G., {et~al.} 2017, \mnras, 465, 843, \dodoi{10.1093/mnras/stw1987}

\bibitem[{{Matsakos} \& {K{\"o}nigl}(2016)}]{matsakos_2016}
{Matsakos}, T., \& {K{\"o}nigl}, A. 2016, \apjl, 820, L8, \dodoi{10.3847/2041-8205/820/1/L8}

\bibitem[{{Mazeh} {et~al.}(2016){Mazeh}, {Holczer}, \& {Faigler}}]{Mazeh_2016}
{Mazeh}, T., {Holczer}, T., \& {Faigler}, S. 2016, \aap, 589, A75, \dodoi{10.1051/0004-6361/201528065}

\bibitem[{{Murray-Clay} {et~al.}(2009){Murray-Clay}, {Chiang}, \& {Murray}}]{MurrayClay_2009}
{Murray-Clay}, R.~A., {Chiang}, E.~I., \& {Murray}, N. 2009, \apj, 693, 23, \dodoi{10.1088/0004-637X/693/1/23}

\bibitem[{{Nortmann} {et~al.}(2018){Nortmann}, {Pall{\'e}}, {Salz}, {Sanz-Forcada}, {Nagel}, {Alonso-Floriano}, {Czesla}, {Yan}, {Chen}, {Snellen}, {Zechmeister}, {Schmitt}, {L{\'o}pez-Puertas}, {Casasayas-Barris}, {Bauer}, {Amado}, {Caballero}, {Dreizler}, {Henning}, {Lamp{\'o}n}, {Montes}, {Molaverdikhani}, {Quirrenbach}, {Reiners}, {Ribas}, {S{\'a}nchez-L{\'o}pez}, {Schneider}, \& {Zapatero Osorio}}]{Nortmann2018}
{Nortmann}, L., {Pall{\'e}}, E., {Salz}, M., {et~al.} 2018, Science, 362, 1388, \dodoi{10.1126/science.aat5348}

\bibitem[{Oklop{\v{c}}i{\'c} \& Hirata(2018)}]{oklopvcic2018new}
Oklop{\v{c}}i{\'c}, A., \& Hirata, C.~M. 2018, The Astrophysical Journal Letters, 855, L11

\bibitem[{Owen(2019)}]{owen2019atmospheric}
Owen, J.~E. 2019, Annual Review of Earth and Planetary Sciences, 47, 67

\bibitem[{{Owen} \& {Jackson}(2012)}]{owen2012}
{Owen}, J.~E., \& {Jackson}, A.~P. 2012, \mnras, 425, 2931, \dodoi{10.1111/j.1365-2966.2012.21481.x}

\bibitem[{{Owen} \& {Lai}(2018)}]{owen_lai_2018}
{Owen}, J.~E., \& {Lai}, D. 2018, \mnras, 479, 5012, \dodoi{10.1093/mnras/sty1760}

\bibitem[{Paragas {et~al.}(2021)Paragas, Vissapragada, Knutson, Oklop{\v{c}}i{\'c}, Chachan, Greklek-McKeon, Dai, Tinyanont, \& Vasisht}]{paragas2021metastable}
Paragas, K., Vissapragada, S., Knutson, H.~A., {et~al.} 2021, The Astrophysical Journal Letters, 909, L10

\bibitem[{{Perez Gonzalez, Jorge}(2023)}]{doi}
{Perez Gonzalez, Jorge}. 2023, Data for TOI-1268 from TESS sectors 15, 21, 22, 41, 48 and 49.,  STScI/MAST, \dodoi{10.17909/MZ3W-V454}

\bibitem[{{Poppenhaeger} \& {Wolk}(2014)}]{Poppenhaeger_2014}
{Poppenhaeger}, K., \& {Wolk}, S.~J. 2014, \aap, 565, L1, \dodoi{10.1051/0004-6361/201423454}

\bibitem[{{Ricker} {et~al.}(2015){Ricker}, {Winn}, {Vanderspek}, {Latham}, {Bakos}, {Bean}, {Berta-Thompson}, {Brown}, {Buchhave}, {Butler}, {Butler}, {Chaplin}, {Charbonneau}, {Christensen-Dalsgaard}, {Clampin}, {Deming}, {Doty}, {De Lee}, {Dressing}, {Dunham}, {Endl}, {Fressin}, {Ge}, {Henning}, {Holman}, {Howard}, {Ida}, {Jenkins}, {Jernigan}, {Johnson}, {Kaltenegger}, {Kawai}, {Kjeldsen}, {Laughlin}, {Levine}, {Lin}, {Lissauer}, {MacQueen}, {Marcy}, {McCullough}, {Morton}, {Narita}, {Paegert}, {Palle}, {Pepe}, {Pepper}, {Quirrenbach}, {Rinehart}, {Sasselov}, {Sato}, {Seager}, {Sozzetti}, {Stassun}, {Sullivan}, {Szentgyorgyi}, {Torres}, {Udry}, \& {Villasenor}}]{TESS2015}
{Ricker}, G.~R., {Winn}, J.~N., {Vanderspek}, R., {et~al.} 2015, Journal of Astronomical Telescopes, Instruments, and Systems, 1, 014003, \dodoi{10.1117/1.JATIS.1.1.014003}

\bibitem[{Salvatier {et~al.}(2016)Salvatier, Wiecki, \& Fonnesbeck}]{salvatier2016probabilistic}
Salvatier, J., Wiecki, T.~V., \& Fonnesbeck, C. 2016, PeerJ Computer Science, 2, e55

\bibitem[{{Salz} {et~al.}(2016){Salz}, {Czesla}, {Schneider}, \& {Schmitt}}]{salz2016}
{Salz}, M., {Czesla}, S., {Schneider}, P.~C., \& {Schmitt}, J.~H.~M.~M. 2016, \aap, 586, A75, \dodoi{10.1051/0004-6361/201526109}

\bibitem[{{Salz} {et~al.}(2015){Salz}, {Schneider}, {Czesla}, \& {Schmitt}}]{Salz_2015}
{Salz}, M., {Schneider}, P.~C., {Czesla}, S., \& {Schmitt}, J.~H.~M.~M. 2015, \aap, 576, A42, \dodoi{10.1051/0004-6361/201425243}

\bibitem[{{Schreyer} {et~al.}(2023){Schreyer}, {Owen}, {Spake}, {Bahroloom}, \& {Di Giampasquale}}]{Schreyer2023}
{Schreyer}, E., {Owen}, J.~E., {Spake}, J.~J., {Bahroloom}, Z., \& {Di Giampasquale}, S. 2023, arXiv e-prints, arXiv:2302.10947, \dodoi{10.48550/arXiv.2302.10947}

\bibitem[{{Schwarz}(1978)}]{annalsofstats}
{Schwarz}, G. 1978, Annals of Statistics, 6, 461

\bibitem[{{Skrutskie} {et~al.}(2006){Skrutskie}, {Cutri}, {Stiening}, {Weinberg}, {Schneider}, {Carpenter}, {Beichman}, {Capps}, {Chester}, {Elias}, {Huchra}, {Liebert}, {Lonsdale}, {Monet}, {Price}, {Seitzer}, {Jarrett}, {Kirkpatrick}, {Gizis}, {Howard}, {Evans}, {Fowler}, {Fullmer}, {Hurt}, {Light}, {Kopan}, {Marsh}, {McCallon}, {Tam}, {Van Dyk}, \& {Wheelock}}]{Skrutskie_2006}
{Skrutskie}, M.~F., {Cutri}, R.~M., {Stiening}, R., {et~al.} 2006, \aj, 131, 1163, \dodoi{10.1086/498708}

\bibitem[{{Spake} {et~al.}(2021){Spake}, {Oklop{\v{c}}i{\'c}}, \& {Hillenbrand}}]{2021AJ....162..284S}
{Spake}, J.~J., {Oklop{\v{c}}i{\'c}}, A., \& {Hillenbrand}, L.~A. 2021, \aj, 162, 284, \dodoi{10.3847/1538-3881/ac178a}

\bibitem[{{Spake} {et~al.}(2018){Spake}, {Sing}, {Evans}, {Oklop{\v{c}}i{\'c}}, {}, {Bourrier}, {Kreidberg}, {Rackham}, {Irwin}, {Ehrenreich}, {Wyttenbach}, {Wakeford}, {Zhou}, {Chubb}, {Nikolov}, {Goyal}, {Henry}, {Williamson}, {Blumenthal}, {Anderson}, {Hellier}, {Charbonneau}, {Udry}, \& {Madhusudhan}}]{spake2018}
{Spake}, J.~J., {Sing}, D.~K., {Evans}, T.~M., {et~al.} 2018, \nat, 557, 68, \dodoi{10.1038/s41586-018-0067-5}

\bibitem[{{Stefansson} {et~al.}(2017){Stefansson}, {Mahadevan}, {Hebb}, {Wisniewski}, {Huehnerhoff}, {Morris}, {Halverson}, {Zhao}, {Wright}, {O'rourke}, {Knutson}, {Hawley}, {Kanodia}, {Li}, {Hagen}, {Liu}, {Beatty}, {Bender}, {Robertson}, {Dembicky}, {Gray}, {Ketzeback}, {McMillan}, \& {Rudyk}}]{2017ApJ...848....9S}
{Stefansson}, G., {Mahadevan}, S., {Hebb}, L., {et~al.} 2017, \apj, 848, 9, \dodoi{10.3847/1538-4357/aa88aa}

\bibitem[{{\v{S}}ubjak {et~al.}(2022){\v{S}}ubjak, Endl, Chaturvedi, Karjalainen, Cochran, Esposito, Gandolfi, Lam, Stassun, {\v{Z}}{\'a}k, {et~al.}}]{vsubjak2022toi}
{\v{S}}ubjak, J., Endl, M., Chaturvedi, P., {et~al.} 2022, arXiv preprint arXiv:2201.13341

\bibitem[{Szab{\'o} \& Kiss(2011)}]{szabo2011short}
Szab{\'o}, G.~M., \& Kiss, L. 2011, The Astrophysical Journal Letters, 727, L44

\bibitem[{{Thorngren} {et~al.}(2023){Thorngren}, {Lee}, \& {Lopez}}]{Thorngren_2023}
{Thorngren}, D.~P., {Lee}, E.~J., \& {Lopez}, E.~D. 2023, \apjl, 945, L36, \dodoi{10.3847/2041-8213/acbd35}

\bibitem[{{Triaud} {et~al.}(2013){Triaud}, {Anderson}, {Collier Cameron}, {Doyle}, {Fumel}, {Gillon}, {Hellier}, {Jehin}, {Lendl}, {Lovis}, {Maxted}, {Pepe}, {Pollacco}, {Queloz}, {S{\'e}gransan}, {Smalley}, {Smith}, {Udry}, {West}, \& {Wheatley}}]{Triaud_2013}
{Triaud}, A.~H.~M.~J., {Anderson}, D.~R., {Collier Cameron}, A., {et~al.} 2013, \aap, 551, A80, \dodoi{10.1051/0004-6361/201220900}

\bibitem[{{Vidal-Madjar} {et~al.}(2003){Vidal-Madjar}, {Lecavelier des Etangs}, {D{\'e}sert}, {Ballester}, {Ferlet}, {H{\'e}brard}, \& {Mayor}}]{vidal_madjar_2003}
{Vidal-Madjar}, A., {Lecavelier des Etangs}, A., {D{\'e}sert}, J.~M., {et~al.} 2003, \nat, 422, 143, \dodoi{10.1038/nature01448}

\bibitem[{Virtanen {et~al.}(2020)Virtanen, Gommers, Oliphant, Haberland, Reddy, Cournapeau, Burovski, Peterson, Weckesser, Bright, {van der Walt}, Brett, Wilson, Millman, Mayorov, Nelson, Jones, Kern, Larson, Carey, Polat, Feng, Moore, {VanderPlas}, Laxalde, Perktold, Cimrman, Henriksen, Quintero, Harris, Archibald, Ribeiro, Pedregosa, {van Mulbregt}, \& {SciPy 1.0 Contributors}}]{2020SciPy-NMeth}
Virtanen, P., Gommers, R., Oliphant, T.~E., {et~al.} 2020, Nature Methods, 17, 261, \dodoi{10.1038/s41592-019-0686-2}

\bibitem[{Vissapragada {et~al.}(2020)Vissapragada, Knutson, Jovanovic, Harada, Oklop{\v{c}}i{\'c}, Eriksen, Mawet, Millar-Blanchaer, Tinyanont, \& Vasisht}]{vissapragada2020constraints}
Vissapragada, S., Knutson, H.~A., Jovanovic, N., {et~al.} 2020, The Astronomical Journal, 159, 278

\bibitem[{Vissapragada {et~al.}(2022)Vissapragada, Knutson, Greklek-McKeon, Oklopcic, Dai, Santos, Jovanovic, Mawet, Millar-Blanchaer, Paragas, {et~al.}}]{vissapragada2022upper}
Vissapragada, S., Knutson, H.~A., Greklek-McKeon, M., {et~al.} 2022, arXiv preprint arXiv:2204.11865

\bibitem[{{Wang} \& {Dai}(2021)}]{Wang_2021}
{Wang}, L., \& {Dai}, F. 2021, \apj, 914, 99, \dodoi{10.3847/1538-4357/abf1ed}

\bibitem[{{Youngblood} {et~al.}(2016){Youngblood}, {France}, {Loyd}, {Linsky}, {Redfield}, {Schneider}, {Wood}, {Brown}, {Froning}, {Miguel}, {Rugheimer}, \& {Walkowicz}}]{Youngblood_2016}
{Youngblood}, A., {France}, K., {Loyd}, R.~O.~P., {et~al.} 2016, \apj, 824, 101, \dodoi{10.3847/0004-637X/824/2/101}

\bibitem[{{Zhang} {et~al.}(2022{\natexlab{a}}){Zhang}, {Cauley}, {Knutson}, {France}, {Kreidberg}, {Oklop{\v{c}}i{\'c}}, {Redfield}, \& {Shkolnik}}]{zhang_2022_dec}
{Zhang}, M., {Cauley}, P.~W., {Knutson}, H.~A., {et~al.} 2022{\natexlab{a}}, \aj, 164, 237, \dodoi{10.3847/1538-3881/ac9675}

\bibitem[{{Zhang} {et~al.}(2023){Zhang}, {Knutson}, {Dai}, {Wang}, {Ricker}, {Schwarz}, {Mann}, \& {Collins}}]{zhang_2023}
{Zhang}, M., {Knutson}, H.~A., {Dai}, F., {et~al.} 2023, \aj, 165, 62, \dodoi{10.3847/1538-3881/aca75b}

\bibitem[{{Zhang} {et~al.}(2022{\natexlab{b}}){Zhang}, {Knutson}, {Wang}, {Dai}, \& {Barrag{\'a}n}}]{zhang2022_feb}
{Zhang}, M., {Knutson}, H.~A., {Wang}, L., {Dai}, F., \& {Barrag{\'a}n}, O. 2022{\natexlab{b}}, \aj, 163, 67, \dodoi{10.3847/1538-3881/ac3fa7}

\end{thebibliography}
\bibliographystyle{aasjournal}

\end{document}